# Revisiting Kunitomo's Sunspot Drawings during 1835-1836 in Japan

**Masashi Fujiyama[1*], Hisashi Hayakawa[2**], Tomoya Iju[3], Toshiki Kawai[1], Shin Toriumi[3], Kenichi Otsuji[4], Katsuya Kondo[1], Yusaku Watanabe[1], Satoshi Nozawa[5], Shinsuke Imada[1]**

(1) Institute for Space-Earth Environmental Research, Nagoya University, 4640814, Nagoya, Japan

(2) Graduate School of Letters, Osaka University, 5600043, Toyonaka, Japan.

(3) National Astronomical Observatory of Japan, 1818588, Mitaka, Japan

(4) Kwasan and Hida Observatories, Kyoto University, Kyoto, Japan

(5) Department of Science, Ibaraki University, Mito, Ibaraki, 310-8512, Japan

\* fujiyama.masashi@j.nagoya-u.jp

\*\* hayakawa@kwasan.kyoto-u.ac.jp

**Abstract**

We revisit the sunspot drawings made by the Japanese astronomer Kunitomo Toubei during 1835-1836 and recount the sunspot group number for each image. There are two series of drawings, preliminary (P, containing 17 days with observations) and summary (S, covering 156 days with observations), all made using brush and ink. S is a compilation of drawings for the period from February 1835, to March 1836. Presently, the P drawings are available only for one month, September 1835; those of other periods have presumably been lost. Another drawing (I) lets us recover the raw group count (RGC) for 25 September 1836, on which the RGC has not been registered in the existing catalogs. We also revise the RGCs from P and S using the Zürich classification and determine that Kunitomo's results tend to yield smaller RGCs than those of other contemporary observers. In addition, we find that Kunitomo's RGCs and spot areas have a correlation (0.71) that is not very different from the contemporary observer Schwabe (0.82). Although Kunitomo's spot areas are much larger than those determined by Schwabe due to skill and instrument limitations, Kunitomo at least captured the growing trend of the spot activity in the early phase of the Solar Cycle 8. We also determine the solar rotation axis to estimate the accurate position (latitude and longitude) of the sunspot groups in Kunitomo's drawings.

# 1 Introduction

Sunspot group counts are widely used to review and reconstruct past solar activity (Vaquero, 2007; Vaquero and Vázquez, 2009; Hathaway, 2010; Clette *et al.*, 2014, 2015, 2016; Clette and Lefèvre, 2014). These data are important not only for reconstructing previous solar activity, but also for providing a basis for theoretical investigations of the solar dynamo (*e.g.* Hotta *et al.*, 2016; Ghizaru *et al.*, 2010; Brown *et al.*, 2011), and prediction of future solar cycles (*e.g.*, Svalgaard *et al.*, 2005; Iijima *et al.*, 2017). Sunspot data span from 1610 and represent one of the longest ongoing scientific records (Owens, 2013; Vaquero *et al.*, 2016).

Rudolf Wolf and his successors systematically collected these records to reconstruct the Wolf sunspot number, which is based both on group and individual spot counts since 1700 (*e.g.*, Waldmeier, 1961). Hoyt and Schatten (1998a, 1998b) conducted further surveys to develop a group sunspot number that extends back to 1610. Recent studies have revisited these two sunspot number time series, added the data, and improved the methodology (*e.g.* Clette *et al.*, 2014; Vaquero *et al.*, 2016; Svalgaard and Schatten, 2016). These attempts are based on fundamental studies of historical sunspot observations and sunspot drawings (*e.g.* Arlt, 2008, 2009a, 2009b; Arlt et al., 2011, 2013, 2016; Carrasco *et al.*, 2015a, 2015b, 2016; Carrasco and Vaquero, 2016; Vaquero, 2007; Vaquero *et al.*, 2007, 2011, 2017; Willis *et al.*, 1996, 2013a, 2013b, 2016; Lockwood *et al.*, 2016a, 2016b, 2017; Usoskin *et al.*, 2015; Svalgaard, 2017; Cliver, 2017; Hayakawa *et al.*, 2018c, 2018e).

In addition to the counts of groups and spots, sunspot areas (Vaquero et al., 2002, 2004; Hathaway *et al.*, 2002), lifetimes, and decay rates (Gesztelyi and Kalman, 1986; Petrovay and van Driel-Gesztelyi, 1997; Petrovay *et al.*, 1999; Henwood *et al.*, 2010; Namekata *et al.*, 2019), and locations (Arlt, 2009; Usoskin *et al.*, 2009; Arlt *et al.*, 2013; Arlt and Weiss, 2014) have also been considered. Based on the datasets of the Royal Greenwich Observatory (RGO), Hathaway *et al.* (2002) found a good correlation between the group sunspot number and the sunspot areas. These original sunspot drawings contribute to our understanding for the space weather studies, due to these multiple information for their areas, positions, morphologies, and evolutions (*e.g.* Silverman and Cliver, 2001; Tsurutani *et al.*, 2003; Cliver and Svalgaard, 2004; Willis *et al.*, 2006, 2009; Vaquero *et al.*, 2008; Cliver and Dietrich, 2013; Shibata *et al.*, 2013; Aulanier *et al.*, 2013; Hayakawa *et al.*, 2016, 2017c, 2018c, 2018d, 2018e, 2019; Lefèvre *et al.*, 2016; Toriumi *et al.*, 2017).

While sunspot observations are geographically concentrated in Europe before the mid 19th century, recent studies have recovered several observations from non-European regions (*e.g.* Denig and McVaug, 2017; Domínguez-Castro *et al*., 2017). In addition to the astral studies (*e.g.* Pankenier, 2013; Morgan, 2017), astronomers in East Asia also pursued solar observations including sunspots (*e.g.* Willis *et al*., 1980, 1996; Yau and Stephenson, 1988; Willis *et al*., 2001, 2005; Hayakawa *et al*., 2015, 2017a, 2017b, 2017c, 2018c, 2018d; Tamazawa *et al*., 2017a). After a cultural contact with European astronomers, East Asian astronomers also began to undertake telescopic observations of sunspots activity (*e.g.* Kanda, 1932; Yamamoto, 1935; Kanda, 1960; Tomita *et al*., 1998; Kubota and Suzuki, 2003; Tamazawa *et al*., 2017b; Hayakawa *et al*., 2018a, 2018b).

Kunitomo Toubei (国友藤兵衛, 1778 – 1840) is one of the most famous early sunspot observers in Japan (Arima, 1932; Yamamoto, 1937; Ota, 2003; Kubota and Suzuki, 2003). He was a gunsmith in the Kunitomo Village (N35°25′, E136°17′) near current Nagahama in Japan, after inheriting his ancestors' position as an official gunsmith. Kunitomo made his own Gregorian reflecting telescope in 1833, inspired by contemporary telescopes from the Netherlands (Yamamoto, 1937; Tomita *et al*., 1998; Ota, 2003; Kubota and Suzuki, 2003). Previously, Kunitomo's sunspot group counts were examined by Kanda (1932), Yamamoto (1935), and Tomita *et al*. (1998). The values obtained by Yamamoto (1935) were adapted by Hoyt and Schatten (1998a, 1998b) and included in the catalog of Vaquero et al. (2016). While these previous studies focused on the summary (S) records of Kunitomo (Kanda, 1932; Yamamoto, 1935; Tomita *et al*., 1998; Hoyt and Schatten, 1998a, 1998b), Kunitomo also left a shorter (one month) preliminary (P) record series of sunspot drawings in September 1835. In addition, there is a single-day observational sheet on 25 September 1836, which was not registered in the group count database (see Vaquero *et al*., 2016). In this article, we revisit Kunitomo's sunspot observations in terms of the sunspot group count and the sunspot area, and compute their ratio and correlation. In addition, a comparison is made between this result and that of other contemporary works.

## 2 Method

In this section, we first describe the source documentation of Kunitomo's sunspot observations and briefly review his telescopes. Then, we examine Kunitomo's method of observing sunspots with his telescope and explain how we derive indices of solar activity such as sunspot group numbers and sunspot areas from his record. We compare the relationship between his group number and the total area of sunspots to evaluate their consistency (see, Hathaway et al., 2002). Finally, we compare the obtained indices of solar activity with those from solar observations by European astronomers and contextualize his observations in our understanding of solar activity in the mid-19th century.

Figure 1. Examples of Kunitomo's sunspot drawing in S (page 6). This image shows the 6th page of S, which contains sunspot drawings obtained on 6, 8, 9, 11, and 12 April 1835.

Figure 2. Kunitomo's sunspot drawing on 25 September 1836 (I). The shaded areas probably indicate the penumbrae of the sunspots.

**2.1 Source Documentation**

An archive of the Kunitomo family (hereafter, Kunitomo Archive) is currently preserved as microfilms as well as printouts in Nagahama City Museum. In this archive, three series of records contain sunspot observations (Ota, 2003). Kunitomo referred to a sunspot as a black spot (黒点, *kokuten*), in the same way as most of the Japanese sunspot records since 851 (Yamamoto, 1960; Hayakawa et al., 2017d, 2018a, 2018b).

There are different types of documents:

Summary (S): *Tenpo Rokunen Shogatsu Kichijitsu Nichigetsu Seigyou Tameshidome*, microfilm 323-5 in Kunitomo Archive, Nagahama City Library.

Preliminary (P): *Tentai Kansoku Zu*, microfilm 324-6 in Kunitomo Archive, Nagahama City Library.

Independent (I): *Tenpo Nananen Sarudoshi Hachigatsu Juugonichi Itsutsudoki-sugi Ukagai*, microfilm 327-9 in Kunitomo Archive, Nagahama City Library.

S is a record of astronomical observations in a scroll form. Its contents are mainly sunspot drawings of the whole solar disk, while some astronomical observations such as drawings of the planet Saturn are also recorded. Kunitomo started astronomical observations in February 1835, using the Gregorian calendar (late 1st month in the 6th year of Tenpo period in a Japanese luni-solar calendar) as the title of S tells. Figure 1 shows an example of sunspot drawings in S. In S, solar disks are drawn in two rows and placed in order of date. Disks in the upper and lower rows show sunspot observations in the morning (around 8 local time, LT) and afternoon (around 16 local time, LT) of a day, respectively. His continuous observations of sunspots were carried out from 3 February 1835 to 24 March 1836 (from the 1st month 6th day in the 6th year to the 2nd month 8th day in the 7th year of Tenpo period). The conversion of Japanese luni-solar calendars to Gregorian ones is based on Uchida (1992). Kunitomo's observational period falls in the ascending phase of the Solar Cycle 8

(see Clette *et al.*, 2014).

P is also a record in scroll form that is totally dedicated to sunspot observations. This document contains sunspot drawings from 2 to 24 September 1835 in the Gregorian calendar (from the 7th month 15th day to the 8th month 3rd day in the 7th year of Tenpo period). Sunspot drawings are presented in the same manner as the document S.

I is a single paper document dedicated to sunspot observations at approximately 8 LT on 25 September 1836 (the 8th month 15th day in the 7th year of Tenpo period). Figure 2 displays a sunspot drawing in I. For this sunspot observation, gray areas are shown around black areas of the sunspots, while these are hardly seen in the drawings in documents S and P.

## 2.2 Telescopes Used in Knitomo's Observations

Here, we briefly review Kunitomo's telescopes. Kunitomo produced a Gregorian reflecting telescope with speculum metal mirrors by himself for the first time in 1833, inspired by contemporary telescopes from the Netherlands (Yamamoto, 1937; Tomita *et al.*, 1998; Ota, 2003; Kubota and Suzuki, 2003). Four of his telescopes, which have a brass mirror tube and pedestal, are currently known. All of them have a Huygens eyepiece (Tomita *et al.*, 1998) a *zonglas* (*sunglass* in Dutch), a colored filter glass to reduce the light passing through, which was imported to Japan until the mid 18th century from the Netherlands (*e.g.*, Zuidelvaart, 1993; Hayakawa *et al.*, 2018a, 2018b).

These telescopes were investigated by Tomita *et al.* (1998) and Ota (2003). The earliest known telescope by Kunitomo is dated 1834 and is currently preserved in the Ueda City Museum. This telescope is 365 mm in length, 62.0 mm in inner diameter, and 68.5 mm in outer diameter for the mirror tube, and has a ball head mount. The first telescope was dedicated to Takashima Fief, although it is not known when this occurred. The second telescope is dated in 1836. It is 334 mm in length, 62.2 mm in inner diameter, and 68.2 mm in outer diameter for the mirror tube, and has an altitude mount. This telescope was dedicated to the chamberlain of Osaka in 1836. The third telescope is undated but considered to have been produced after the first two telescopes since its prop is improved to be movable. This telescope is 348 mm in length, 61.6 mm in inner diameter, and 69.0 mm in outer diameter for the mirror tube, and has an alt-azimuth mount. It was dedicated to the Hikone Fief in 1842. The fourth telescope is preserved in the household of Kunitomo family without an inscription of dating. Nevertheless, it has the same improvement in its prop, similar to the third telescope. Only its length is known to be 345 mm. The optical system of Kunitomo's telescope was studied by Tomita et al. (1998). The earliest telescope has a 60 mm primary mirror, an 11.1 mm secondary mirror, and a focal length of 3,520 mm. Therefore, this telescope with the Huygens eyepiece, which has a focal length of 53.4 mm (Tomita et al., 1998), obtains a magnification of about 66 times.

While it is not known which telescope was used in his sunspot observations, we can exclude three of the telescopes except for the first one, as they were built after the start of his observation on 2 March 1835. At the same time, we also need to be aware that the production of the earliest telescope by Kunitomo was in 1832. Considering this, it is also possible that none of the four telescopes is the earliest one produced by Kunitomo, and the telescope in use in his sunspot observations may not have been preserved.

**2.3 Kunitomo's Method of Drawing Sunspots**

Here, we describe Kunitomo's method of drawing sunspots with his Gregorian telescope. Tomita *et al*. (1998) pointed out that Kunitomo Toubei probably drew sunspots on a Japanese paper using a writing brush and Chinese ink and hence it may not be possible to distinguish umbrae and penumbrae in a sunspot due to the run of ink on the paper. We confirm that the sunspots have no clear penumbrae in documents S and P, whereas the document I shows dark spots with dim structures, which possibly indicate sunspots with penumbrae (Yamamoto, 1937). We note that the disks in I have a larger size than the disks in S and P. Hence, it is suggested that the absence of penumbrae in S and P stems not from the astronomical instrument used, but from the drawing method.

The document I has a sunspot drawing of the whole solar disk with a caption, which has a sentence detailing the daily motion of sunspots. From this part of the caption and the sunspot drawing, we can roughly determine the directions of the solar equatorial line and of the solar differential rotation. For a large sunspot group in the first quadrant of the sketch in Figure 2, which may be identified as an F-type group in the Zürich classification (Waldmeier, 1947; Kiepenheuer, 1953), we determine the major axis of this sunspot group. Since the major axis makes an inclination angle of a few degrees with the solar equator (Joy's law; Hale *et al*., 1919), we find that the solar equatorial line is drawn from the bottom left to top right in the solar circle in the document I. The identified direction of the equatorial line is consistent with the direction of the daily sunspot motion described in the caption. Furthermore, from a comparison between the sunspot drawings in successive days in documents S and P, we find that the solar equatorial line is again drawn approximately from the left to right or from the bottom left to the top right for most of the sunspot drawings made in the morning, and from the top to bottom or from the top left to the bottom right for most of those made in the afternoon.

The daily motion of sunspots indicated in S suggests that Kunitomo watched sunspots through his telescope with the *zonglas*, colored filter glass, and drew them on a paper using the drawing tools previously mentioned. Tomita *et al.* (1998) also presented a similar suggestion. In other words, Kunitomo probably did not use a projection technique for these sunspot observations. Therefore, the data related to the sunspot position of Kunitomo's observations have less accuracy than those from sunspot observations based on the projection method. In fact, we found that sunspot groups occasionally changed their latitudinal location by more than 10 degrees in successive drawings. We therefore consider that Kunitomo's sunspot drawings have the sufficient quality to estimate the daily number of sunspot groups but not to locate the positions of the spots. This is the reason why we evaluated the uncertainty of sunspot locations in Kunitomo's sunspot drawings in comparison with those in Schwabe's drawings (Arlt *et al.*, 2013).

**2.4 Identification of Sunspot Group from Kunitomo's Solar Observations**

We interpret sunspot groups in the sunspot drawings of S and P, based on the Zürich classification as defined in Kiepenheuer (1953). Before grouping the sunspots, we determined the direction of the solar rotation for each sunspot drawing because Kunitomo did not indicate this information. Sunspots with relatively large sizes such as D, E, and F-type groups in the Zürich classification system have longer lifetimes and their longitudinal extension is larger than their latitudinal extension. We infer the position of the solar equator from the major axis of relatively large sunspots in the sunspot drawings. The fact that the relative position between sunspot groups is maintained for several days helps us to determine the direction of the solar rotation from a comparison between sunspot drawings on successive days.

For sunspot drawings for which the solar equator was found, we classified them into different categories based on the Zürich classification system with the following criteria: (1) If a small sunspot is distant from a larger one by five degrees or more in the latitudinal direction, it is classified as another group. (2) If a small sunspot appears in the vicinity of a larger one in the longitudinal direction, it is interpreted as part of the larger developing sunspot group. After considering the classification of the sunspot groups, we counted the daily number. Figure 3 shows examples of the sunspot classification with our criteria for drawings made in the afternoon on 6 and 7 September 1835. It is found from a comparison between these drawings that the direction of solar rotation is from the top left to the bottom right. We identified four sunspot groups in both of the drawings.

Figure 3. Classification examples of sunspot groups in Kunitomo's sunspot drawings made in the afternoon on 6 and 7 September 1835. Left and right panels display the 6 and 7 September drawings, respectively. Each of the sunspot groups is enclosed by a red ellipse. In the 6 September drawing, dark features without a red ellipse are characters for the identification of sunspots.

**2.5 Scaling the Sunspot Area**

In order to scale sunspot areas in these drawings, we performed the following procedures. Here, we use document S, which contains the largest number of sunspot drawings in the documents of Kunitomo's solar observation. Document S is photographed and digitized with an 8 bit grayscale level, and the brightness is expressed from 0 to 255 ($2^8$-1). Firstly, we cut out solar disks from each image. Next, we apply an elliptical fitting to these solar disks and reform them accordingly so that the ellipse becomes a circle, after subtracting the background by the method in the Section 2.2 of Toriumi *et al.* (2014). Then, we determine the center position and radius of the solar disks.

Secondly, we make a binary map of the background for each disk in order to extract sunspots. In this process, we use the intensity threshold of 255 − 5σ for the background, where σ means the standard deviation of the brightness for the reformed images, and is given as 11.9. Using the binary map, sunspots are extracted from the reformed images.

Thirdly, we project the extracted sunspots to a plane using a sinusoidal projection in order to measure an area for the spots. For this projection, the area per pixel is the same everywhere and is set as $1.474 \times 10^6$ km$^2 \approx 0.484$ msh in this study[1]. After performing this processing, we examine the total sunspot area on each disk.

**2.6. Determination of the Solar Rotation Axis**

We determined the solar rotation axis to estimate the accurate position (latitude and longitude) of the sunspot groups in Kunitomo's drawings. Determining the rotation axis is not straightforward, because Kunitomo sketched the solar disk not by projection but by direct vision through a filter. The time unit system at that time, which had roughly one-hour variation back and forth (Uchida, 1992; Soma *et al.*, 2004), is also one of the sources of the uncertainty. Further, the longest series S does not come from the original observational sheets but is a summary from the original sheets, as shown in the philological discussions in this article. Therefore, we evaluated the uncertainty of locations of sunspot groups in Kunitomo's sunspot drawings in comparison with those of Schwabe (Arlt *et al.*, 2013), who is known as a contemporary backbone observer for modern sunspot number reconstruction (Svalgaard and Schatten, 2016).

---

[1] Millionths of Solar Hemisphere: The unit of sunspot area generally used when the size of sunspots is mentioned.

## 3 Results

**3.1 Two Manuscripts for the Same Observational Dates in September 1835.**

P and S provide us with sunspot observations on the same day and time, which implies that one of them is the original observation sheet while the other is its compilation. Figure 4 shows a comparison of sample drawings from the same observational time from P and S. Yamamoto (1935) considers P to be the original sheet and S as the copy, probably due to their time span in his Note 2. The arrangement of the observational time of the sunspots seems to support his idea. In P and S, the circles in the upper row represent observations in the morning and those in the lower row highlight observations in the afternoon. We suggest two reasons why P is an original sheet. Firstly, P has three blank disks in the lower row, which are drawn but observations are not shown (13, 15, and 17 September 1835), while S has days without disks in the lower row. Secondly, on 4 September, P has the observation in the morning drawn in the lower row, where the afternoon observations should be drawn, while S has that observation drawn in the upper as usual. Therefore, it seems that Kunitomo drew the observation in the wrong position in P, and he rewrote that observation in the correct position when he compiled S. For these reasons, we conclude that P is a part of the original observation sheet and S is its compilation. Unfortunately, P covers only 17 days in September 1835, while S covers 156 days in total[2]. This means that the majority of these original sheets are currently lost or at least not available. Yamamoto (1935) suggests that there were also original sheets of sunspot drawings on 3, 9, 11, and 16 April 1835, but they are not found or confirmed in the microfilms currently available at the Nagahama City Museum.

**Figure 4.** Examples of Kunitomo's sunspot drawings in (Left) P (page 2) and (Right) S (page 22). The observational time of these drawings is approximately 16 LT on 6 September 1835. We note that the three black spots on the left-hand side of P are not present in S.

---

[2] Note that Kunitomo actually provides disks for 157 days, but one of them is without observational record due to rain (3 March 1835)

### 3.2. Raw Group Count to Add

The sunspot record for 25 September 1836, is newly recovered from I. The group number for this day was not registered in any of the previous studies. We hereby classify the sunspots into different classes of the Zürich classification system and obtain the raw group count (RGC) of seven from this drawing. Figure 5a shows how we classify sunspots into seven groups in Kunitomo's sunspot drawing (I) using Zürich classification. It is noted that this sunspot drawing is much more detailed than the other drawings in S and P and shows black spots surrounded by gray zones, which presumably correspond to umbrae with penumbrae (see Section 2.3). Yamamoto (1937) considered that it was in September 1836 when Kunitomo first noticed the umbrae and penumbrae.

Figure 5. (left) The sunspot drawing on 25 September 1836 by Kunitomo (08:00 LT ±1 h) and (right) reconstructed umbral distribution on 27 September 1836 by Schwabe (15:30 LT), based on the data of umbral area and umbral position in Arlt et al. (2013). In both of panels, each of sunspot group is enclosed by a red ellipse and alphabets above ellipse. Letters above the ellipses indicate the same group. Seven and five of sunspot groups are identified in the left and right panels, respectively. Sets of the a1 and a2 groups and of the b1 and b2 groups in the Kunitomo's drawing are classified into the a and b groups in the Schwabe's observation, respectively. Note that this sunspot drawing by Kunitomo is rather detailed and includes penumbrae differing from his other drawings in P and S series. The solar rotation axis is tilted up to about 43 degrees counter clockwise from the vertical line in both of panels. Nevertheless, the sunspot groups in these drawings show a considerable uncertainty on their latitude love location and hence the reliability of sunspot positions are subjected to the discussions in the Section 4.

We contextualize Kunitomo's observations in the latest archive of raw group counts by known sunspot observers in the database of Sunspot Index and Long-term Solar Observations (SILSO) (Vaquero et al., 2016). This database shows a gap of three days without known observations within the records on 22 September by Schwabe, on 26 September by Stark, and 27 September by Schwabe (see, Figure 5 right panel). Figure 6 shows the newly obtained Kunitomo's RGC on 25 September and the RGCs by known observers in the epoch between 1 September and 21 October 1836. We consider this observation by Kunitomo to be in good agreement with the contemporary observations by Schwabe, Stark, and Hussey, in terms of their RGCs. We also examine the distribution of sunspots in the Kunitomo's sunspot drawing (I) and that in a contemporary observation. Figure 5 shows positions of sunspot groups in Kunitomo's observation (I) on 25 September and those from Schwabe's observation on 27 September. The latter is reconstructed based on the data of sunspot area and sunspot position in Arlt *et al.* (2013). Schwabe identified five sunspot groups on 27 September 1836. We note that sunspots on the left panel of Figure 5 seem to have a larger size than their actual one and their latitudinal locations have less accuracy due to Kunitomo's methodology for the sunspot observation as mentioned in the Section 2.3. We will discuss the difference between sunspot area of Kunitomo's drawings and umbral area of Schwabe's drawings in Section 4. Nevertheless, Figure 5 shows Kunitomo and Schwabe recorded the same sunspot groups in their sunspot drawings, although they had a different drawing technique. We find four of the same sunspot groups, which are indicated with letters, in both of observations when comparing the left and right panels in Figure 5 regardless the facts that the solar equatorial line is drawn from the bottom left to the top right in Kunitomo's observation (I) and that Schwabe made his sunspot drawing two days later Kunitomo's observations. It is noted that sets of the a1 and a2 groups and of the b1 and b2 groups in the Kunitomo's drawing are classified into the a and b groups in the Schwabe's observation, respectively, and this difference in grouping stems from the difference between Zürich classification (Kiepenheuer 1953) and Schwabe's classification. From this example, we consider the Kunitomo's sunspot drawing on 25 September 1836 to be reliable in terms of the relative distribution of sunspot groups.

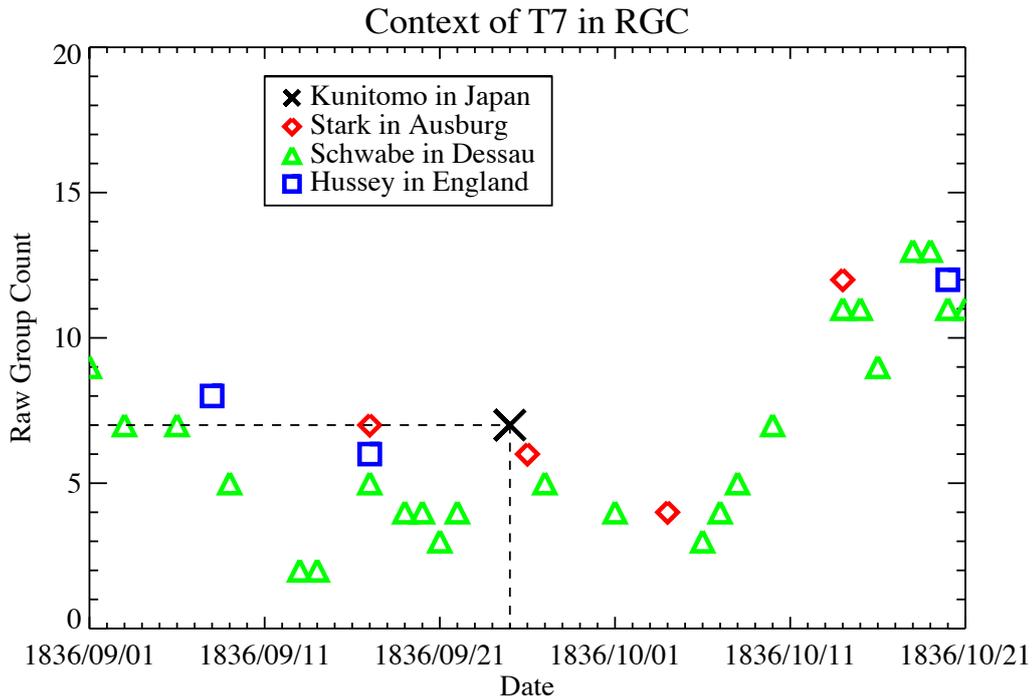

Figure 6. The context of RGC in I in function of time. Black crosses correspond to Kunitomo's observation (I) on 25 September 1836. Other symbols show RGCs of known observers in this epoch.

### 3.3. RGC to Remove or Revise

We revise some RGCs based on the description of the microfilm. We recounted all the sunspot groups in Kunitomo's observations according to Zürich classification (Waldmeier, 1947; Kiepenheuer, 1953) to make these data consistently comparable with modern observational data (see also Clette *et al.*, 2014).

In this process, we have also revised some dates of Kunitomo's observations and resolved data with discrepancies in previous studies. Firstly, we examine the discrepancy in RGCs for five days, *i.e.* 11, 12, 13, 17, and 19 May 1835 in the previous studies. While Kanda (1932) and Yamamoto (1935) register these observations, Tomita *et al.* (1998) omitted them. We carefully resurveyed the microfilms in Nagahama City Museum and found these observations in the 11th frame, while they are not present in the printouts stored in the same museum. We therefore include these five observations in our list.

Secondly, we remove the observation on 10 February 1835 since this is not a spotless day, but a day without observations. Kunitomo states he did not observe during this period (S, page 4). This means that he did not make observations on this day and hence we cannot determine the real group count. Schwabe reports two sunspot groups on the same day (*e.g.* Vaquero *et al*., 2016).

Thirdly, we revise the observation date for a sunspot drawing in the 32nd frame of S from 26 January 1836 to 28 December 1835, based on its original description. The description is written as the "9th day in the same month" in the section of the 11th month, while it appears after the "17th day in the same month". Kanda (1932), Yamamoto (1935), and Tomita *et al*. (1998) interpreted this as the 9th day in the "12th month", putting emphasis on the chronological order from right to left. While we do not discard this possibility, one should also consider the possibility that S is compiled from the original observational reports (*e.g.* P). In this case, it is not too surprising to expect chronological disorder in S. After all, we need to note that the observational date of this record is doubtful.

Lastly, we note that the afternoon observation on 5 September 1835 is included in Yamamoto (1935), while he considers that this observation is also found in S. However, S does not have the observation of this date. These revised RGCs are plotted in Figure 7, in comparison with those by other observers in this epoch based on Vaquero *et al*. (2016).

Lists of Kunitomo's revised RGCs from S and P are given in Tables 1 and 2. Table 1 contains RGCs from this study with RGCs from Kanda (1932), Yamamoto (1935), and Tomita *et al*. (1998) for S. In this table, month, day, and time are given as a local mean time at the Kunitomo's observational site, *i.e.* the Kunitomo Village (N35°25′, E136°17′) in Japan. Table 1 and Figure 7 confirm that our revised RGCs for Kunitomo's observations is consistent with the RGC values from the previous studies in its trend. Table 2 provides RGCs from P and I.

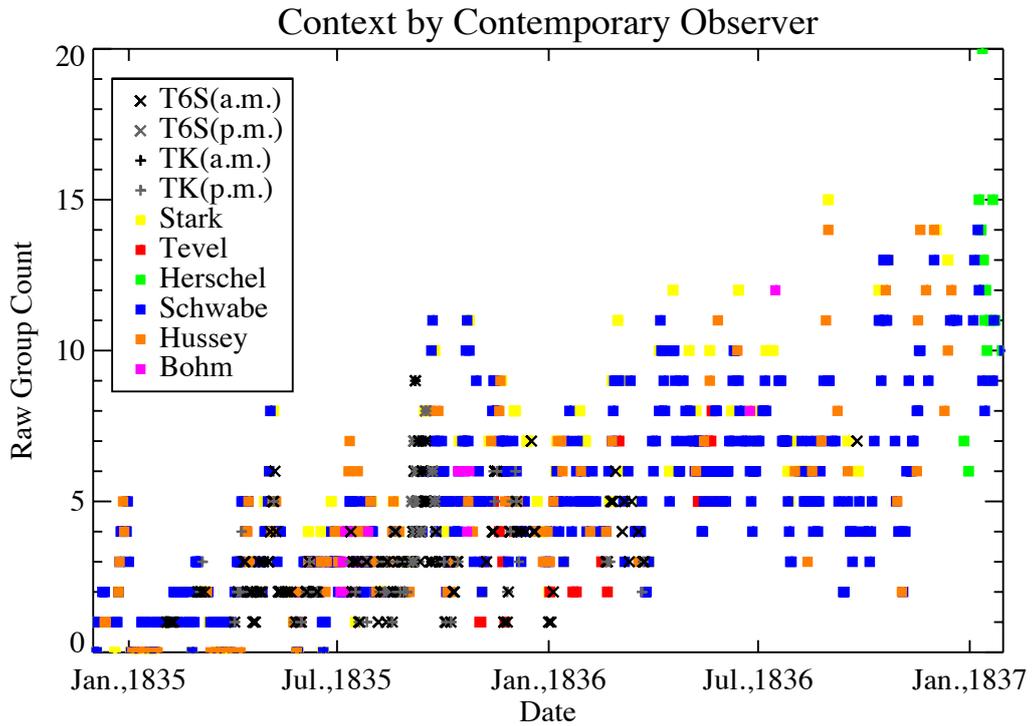

Figure 7 The context of Kunitomo's RGCs in comparison with those by other contemporary observers.

## 4. Discussion

Figure 7 shows that Kunitomo's RGC is consistent with other observers' RGCs in its trend but generally smaller than those of other observers derived from the archive of RGCs of known sunspot observers (Vaquero *et al.*, 2016). This difference is clearly shown in the comparison between the solar observations performed on the same days.

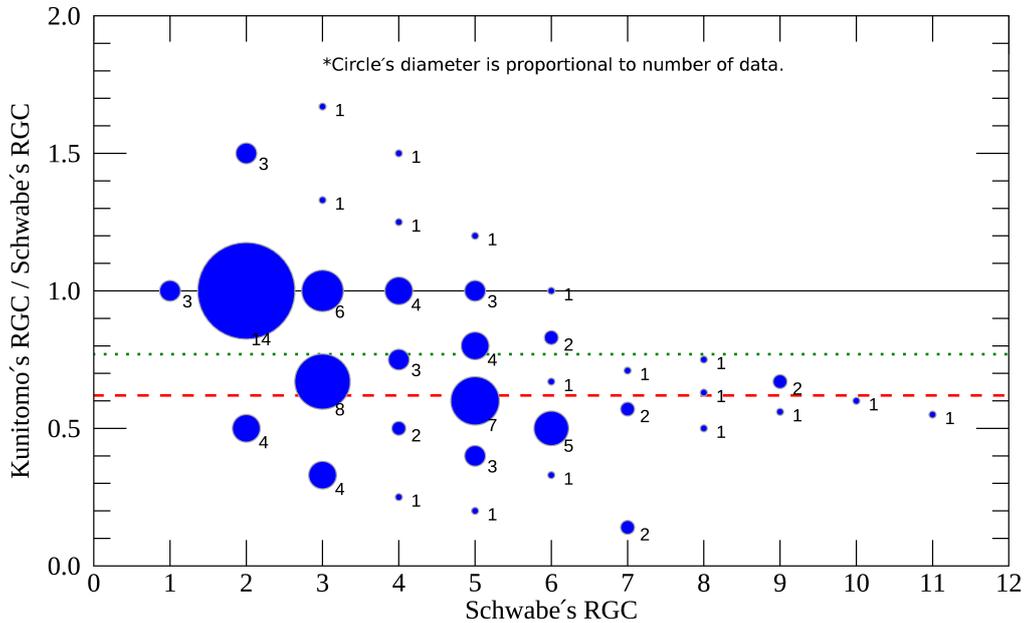

Figure 8 Ratio of Kunitomo's RGC series to that of Schwabe, shown as a function of Schwabe's RGC. The diameters of the filled circles and the numbers noted in their vicinity exhibit the number of data for each data point. The black solid, green dotted, and red dashed lines represent a ratio of 1.0, mean ratios of 0.77 for all data, and of 0.62 for data from Schwabe's RGC $\geqq$ 5, respectively.

Figure 8 shows the ratio of Kunitomo's RGC to Schwabe's RGC as a function of Schwabe's RGC. This figure tells us that the majority of data points have a ratio of 1.0 or less in this comparison. The underestimation of Kunitomo's RGC is very clear for data with Schwabe's RGC $\geqq$ 5; this has a mean ratio with a standard deviation of 0.62±0.23. These numbers are 0.77±0.31 for all data as shown in Figure 8. It is of interest why this underestimation was caused in Kunitomo's RGC. Examinations of an archive of RGCs by Vaquero *et al.* (2016) and two sunspot drawings by Schwabe published in Arlt (2011) and Arlt *et al.* (2013) tell us that these RGCs by Schwabe in the archive probably equal the number of sunspot groups classified by Schwabe himself. Schwabe considered a sunspot group as a cluster of sunspots, which was isolated and not connected to other clusters of spots or cloud-like feature (Svalgaard and Schatten, 2016). Nevertheless, we consider that a difference between Schwabe's and Zürich classifications does not explain all the reasons why Kunitomo's RGC became two thirds as much as those of Schwabe's RGC, because a group counting with the Zürich classification in the Schwabe's drawings give an RGC close to Schwabe's one.

Therefore, it is more plausible than the underestimation of Kunitomo's RGC might have been caused by Schwabe's superior instruments and techniques in comparison to Kunitomo's. Schwabe used two Keplerian telescopes, *i.e.* a 3.5- and six-feet telescopes, to observe sunspots since 1826 and 1829, respectively (Arlt, 2011). He reduced the apertures to 1.75 inches for the 3.5-feet telescope and 2.5 inches for the six-feet telescope for solar observations (Johnson, 1857) and watched sunspots through his telescopes with a magnification of 45 – 96 times and dimming glasses (Johnson, 1857; Arlt, 2011). The Gregorian reflectors operated by Kunitomo had the disadvantage of thermal air convection within a mirror tube, which could have affected the quality of the solar image, in comparison to the Keplerian refractors used by Schwabe. Kunitomo observed sunspots at approximately 8 LT and/or 16 LT, while Schwabe took sunspot drawings at noon (Arlt, 2011). A low solar altitude also led to less ideal observational conditions. Therefore, we consider that Schwabe overlooked fewer spots than Kunitomo's S series, thank to his observational instruments.

Additionally, we must note the possibility that Kunitomo omitted some spots when he compiled S from P as shown in Figure 4. From a comparison between RGCs on the same days in Tables 1 and 2, one may note that 16 of the 32 drawings in P have smaller numbers than S. We notice that occasionally, S does not have some spots drawn in P.

Although Kunitomo's sunspot observation suffered from instrument restrictions and methods, as mentioned above, the series of Kunitomo's drawings is still important, since he was the second most active observer during 1835 -- 1836 after Schwabe. This may contribute to strengthen the known temporal variation of RGC obtained by other contemporary observers in further detail.[3]

It is known that the monthly mean value of the total sunspot area and the group sunspot number have a good correlation with each other. According to Hathaway et al. (2002), the correlation coefficient is 0.988. Here, we evaluate the reliability of sunspot areas in Kunitomo's drawings by examining the aforementioned relationship. It should be noted that Hathaway et al. (2002) used the group sunspot number of Hoyt and Schatten (1998a, 1998b) derived from the RGCs of various observers, while we use the RGCs derived from Kunitomo's document S. We calculate the correlation between the monthly mean total sunspot areas and the monthly mean RGCs of Kunitomo's observations. We also calculate Schwabe's observations in the same time period as

---

[3] From 1 January 1835 to 31 December 1836, Schwabe observed during 422 days, whereas Kunitomo did so during 156 days. Stark, Hussey, Tevel, Bohm and Herschel performed observations for 147, 112, 25, 20, and 1 days, respectively. See Vaquero et al. (2016) for the number of days with observations for other sunspot observers.

Kunitomo. We obtained Schwabe's RGC from Vaquero *et al.* (2016) and the umbral areas from Senthamizh Pavai et al. (2015). In this analysis, we define the sunspot area as the sum of the umbrae and penumbrae in Kunitomo's drawings since Kunitomo does not distinguish umbrae and penumbrae except for I, while only the umbral areas are used for Schwabe's observations.

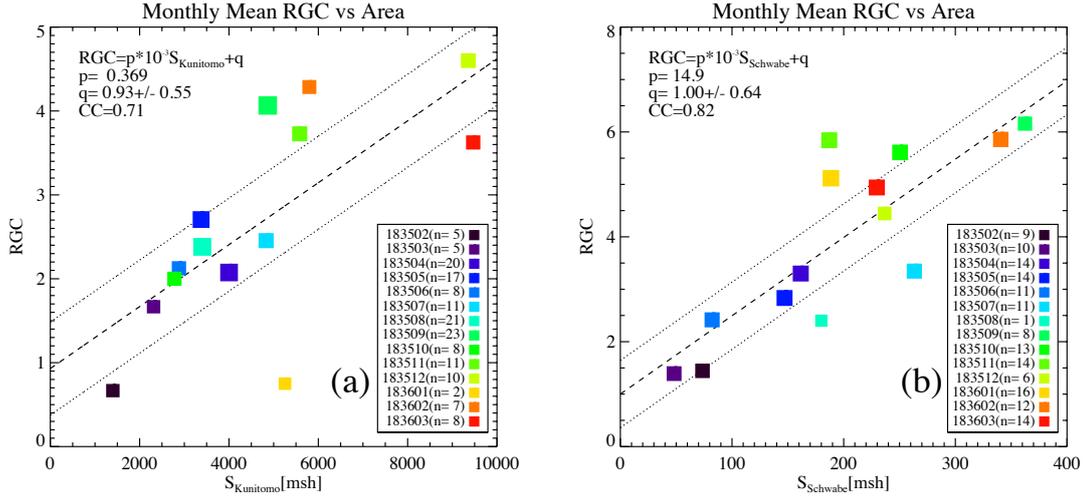

Figure 9. The relationship between the monthly mean RGCs and the sunspot areas derived from (a) Kunitomo and (b) Schwabe. Symbol sizes indicate the amount of data and colors show the month. Dashed and dotted lines indicate the linear fitting line and 1-σ uncertainty in linear fitting, respectively. $S_{Kunitomo}$ has a weaker correlation coefficient (CC) for RGC with a slightly looser fit than $S_{schwabe}$ does.

Figure 9 shows that the sunspot areas from Kunitomo's observations ($S_{Kunitomo}$) has a slightly worse correlation than that those from Schwabe's observations ($S_{Schwabe}$). We obtain a correlation coefficient of 0.71 for Kunitomo and 0.82 for Schwabe. Considering that we only have 14 data points in Figure 9, the obtained correlation coefficient of 0.71 for Kunitomo is relatively high and thus his spot areas may reflect the actual solar conditions to some extent.

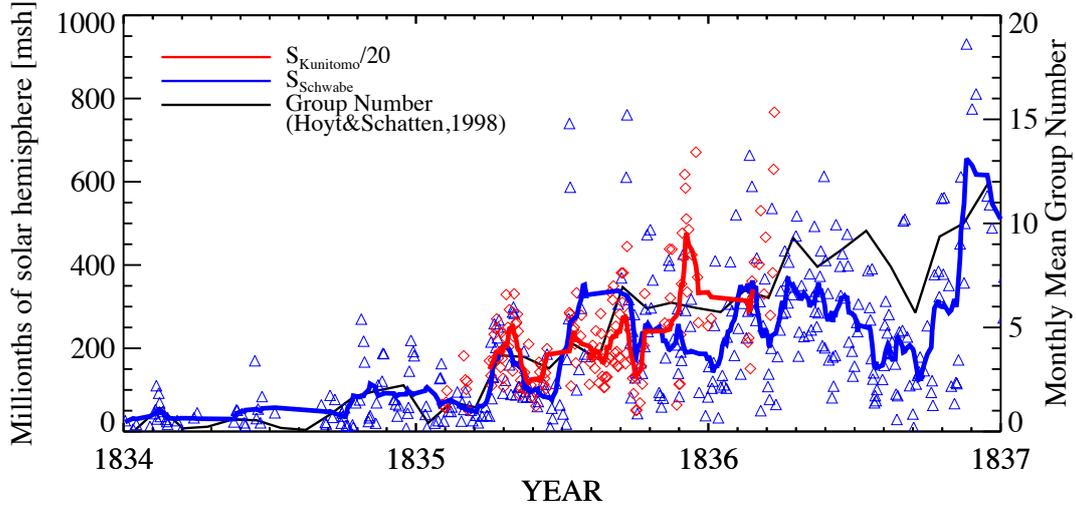

Figure 10: The temporal variations of sunspot areas derived from the drawings of two observers. Left vertical axis indicates the spot group area in msh. Red diamond symbols represent $0.05 \times S_{Kunitomo}$ and blue triangles denote $S_{Schwabe}$. Red and blue solid lines show running averages in 10 plots for $0.05 \times S_{Kunitomo}$ and $S_{Schwabe}$, respectively. The vertical axis on the right is for the monthly mean value of the group numbers from Hoyt & Schatten (1998), shown in this diagram as the black solid line.

Figure 10 shows the temporal variations of the sunspot areas from Kunitomo's and Schwabe's observations in comparison to the monthly mean of the RGCs. This figure tells us that the sunspot areas derived from Kunitomo's observations ($S_{Kunitomo}$) are about 20 times larger than the umbral areas from Schwabe's sunspot drawings ($S_{Schwabe}$), but that the variation trends of $S_{Kunitomo}$ and $S_{Schwabe}$ are similar to each other. Considering the ratio of umbra to penumbra is generally about 5 times (Brandt et al., 1990), it should be noted that the difference is considerably large and that Kunitomo's observations may be relatively exaggerated. We have two hypotheses to explain this discrepancy. One is that Kunitomo used ink and brush to draw sunspots and hence each sunspot is occasionally connected. The other is related to the blur of each sunspot drawn in ink. The large discrepancy between the areas of Kunitomo's sunspot drawings produced with brush and ink and those of Schwabe causes us to speculate that the spot areas of other observers who similarly use brush and ink such as Iwahashi Zenbei (Hayakawa *et al.*, 2018a, 2018b) also have comparable uncertainties.

However, we still find chronological increases of both the total sunspot area and group number in Kunitomo's observations in Figures 7 and 10. Kunitomo's observations occur in the ascending phase of the Solar Cycle 8, from the minimum of 1833 to the maximum of 1837 (*e.g.* Clette *et al.*, 2014; Svalgaard and Schatten, 2016). Therefore, although Kunitomo's drawings are less reliable than those of Schwabe, we consider that he tried to depict the spot areas for as long as his skills and instruments permitted, and as a result, he captured at least the growing trend of the solar activity.

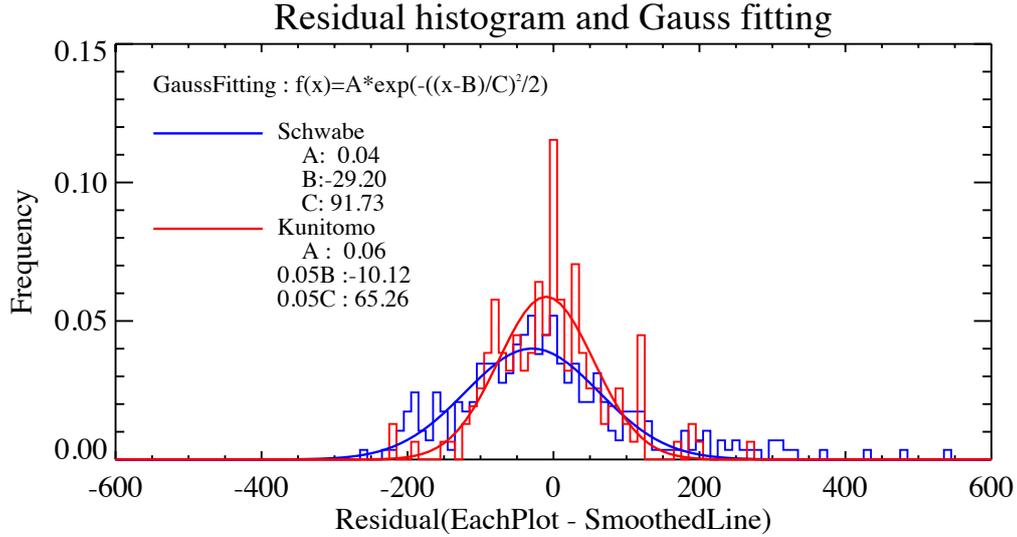

Figure 11: Residuals of the sunspot area evolutions derived from Figure 10 after subtracting a smooth curve (from 3 February 1835 to 24 March 1836 corresponding to Kunitomo's observational period). The histogram is generated using a bin size of 10. The continuous lines are Gaussian fittings. A, B, and C are the height, the center, and the width (standard deviation) of the Gaussians, respectively.

In the same way as Figure 10 shows the time evolution of sunspot areas, it would be helpful to consider the contribution of the random component for the sunspot areas in Kunitomo's sunspot drawings based on how well the residuals fit the normal distribution. We calculated the residuals subtracting the smoothing lines from data points in Figure 10. We generated the histogram with a bin size set as 10. We then made a Gauss fitting to the obtained histogram as shown in Figure 11. As a result, we obtained a smaller value of C (σ ≈ random component) in Kunitomo. As we find Kunitomo's sunspot area is 20 times as large as Schwabe's umbral area, we multiply this factor (x20) to C in Kunitomo. Then, we conclude that the contribution of the random component is considered larger in Kunitomo's observations. Further, we obtained the relatively large negative values of B (centroid) for both of Kunitomo and Schwabe. This is probably because of the distortion in the profiles. We have checked that the average values are almost 0.

We also try to estimate the solar rotation axis in Kunitomo's sunspot drawings. As we mentioned in Section 2.6, it is not straightforward to determine the locations of sunspot groups or reconstructing the butterfly diagram from Kunitomo's sunspot drawings. We try to evaluate the uncertainty of locations of sunspot groups in Kunitomo's sunspot drawings in comparison with those of Schwabe (Arlt et al., 2013), a contemporary backbone sunspot observer in Clette *et al.* (2014) and Svalgaard and Schatten (2016). We use the information of sunspot locations obtained from the Schwabe's sunspot observations (Alrt et al., 2013) as the reference to calibrate the solar rotation axis direction of Kunitomo's sketches. We calibrate the sunspot locations of Kunitomo's drawings by the method discussed in Section 2.6. Specifically, we derotated the solar images so that the rotation axes are aligned to the vertical direction, using the following steps. First of all, we calculated the theoretical tilt angle of the solar rotation axis for each Kunitomo's sunspot drawings according to the observed timings. The tilt angle is obtained by following equation.

$$\tan\theta = \frac{\cos\varphi \sin t}{\sin\varphi \cos\delta - \cos\varphi \sin\delta \cos t} \ .$$

In this formula, $\theta$ is the tilt angle of the solar image, $\varphi$ is the latitude of the sunspot location, $t$ and $\delta$ are the hour angle and the declination of the Sun, respectively. We tentatively derotated the solar images using this tilt angle and obtained the preliminary corrected images. Nevertheless, we need to note that the observation timings in Kunitomo's observations are not so accurate, due to the contemporary time system (Uchida, 1992). Therefore, we performed a further adjustment on the images using Schwabe's sunspot drawings as reference. We rotated the preliminary corrected images step by step and obtained the rotation angle that best matches Kunitomo's and the reference Schwabe's drawings. Finally, we derived the most likely direction of the solar rotation axis for Kunitomo's drawings. Figure 12a shows the histogram of the measured sunspot latitude distribution for the three datasets: Kunitomo's preliminary corrected drawings, finally calibrated drawings, and Schwabe's reference drawings. The horizontal axis shows the observed sunspot latitudes, and the vertical axis shows the occurrence of the sunspots normalized by the observed total number of sunspots. The sunspot distributions of preliminary-corrected Kunitomo's observations and finally-calibrated Kunitomo's and Schwabe's observations are plotted in red, blue, and black solid lines, respectively. In Figure 12a, we can see two peaks for Schwabe (black) and calibrated Kunitomo (blue), which is a result consistent to a butterfly diagram. Please note that our results are derived from around one year data. On the other hand, when we estimate the solar rotation axis using the time information only, the histogram of sunspot latitude (red) shows a single peak around the equator. Figure 12b shows the histogram of an additional rotation angle required when matching the preliminary-corrected Kunitomo's drawing and Schwabe's observations. Because the standard deviation of the rotation angle (σ) is 18.4 degrees, there might be ≈ 20 degrees uncertainty for the rotation axis in Kunitomo's drawings. This uncertainty might arise from the imprecision of observation timing, mispositioning of sketched sunspots (due to Kunitomo's drawing ability and atmospheric seeing). It is also noted that the peak of the histogram offsets to about -10 degree. This might imply that the time in Kuniomo's observations had a systematic delay, the timing information has been written on the sketch before the sunspot drawing started, Kunitomo had a tendency to tilt his head to the left when observing the Sun through the telescope, or the sketching paper tilted counter-clockwise slightly.

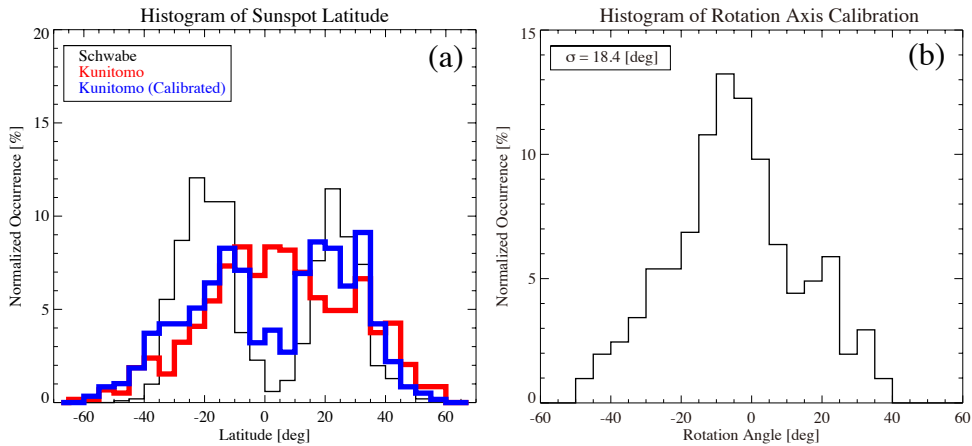

Figure 12: (a) Histograms of sunspot latitudes. The sunspot distributions of preliminary-corrected Kunitomo's observations and finally-calibrated Kunitomo's and Schwabe's observations are plotted in red, blue, and black solid lines, respectively. (b) Histogram of an additional rotation angle required when matching the preliminary corrected Kunitomo's drawing and Schwabe's observations.

## 6. Conclusion

In this article, we revisited Kunitomo's sunspot drawings to obtain the RGCs as well as the total spot areas. Firstly, we clarified that Kunitomo left three kinds of sunspot drawings and reviewed his Gregorian reflecting telescopes. We then grouped the sunspots in his drawings by following the Zürich classification and derived daily RGCs. The RGC for 25 September 1836, was obtained for the first time in this study. We also updated RGCs for other days according to the Zürich classification, which is consistent with the modern classification method. By comparing the obtained RGCs with those of contemporary observers, we found that Kunitomo results tend to provide a smaller number for the RGCs. We measured the sunspot areas from Kunitomo's drawings and confirmed that the areas correlate with the RGCs. Although Kunitomo's spot area is generally much larger and thus less reliable than that of Schwabe, we found that both Kunitomo's RGCs and the spot areas increase as time progresses, which is fairly reasonable because his observation period is situated in the ascending phase of the Solar Cycle 8. Therefore, we conclude that Kunitomo's sunspot drawings can contribute to a reconstruction of past solar activity despite its smaller RGCs and large sunspot areas. On the other hand, it appeared that the sunspot locations of Kunitomo's observation are not necessarily reliable enough due to his observational method, error value of observational time, and handwritten copy process to the summary sheets (S) from the original sheet. Using Schwabe's sunspot observations we estimate the solar rotation axis, and we find an uncertainty $\approx$ 20 degrees for the rotation axis in the Kunitomo's drawings.

**Acknowledgements**

We thank Masanobu Kunitomo for providing permissions to use the figure and data for Kunitomo's observation, and the Nagahama City Museum for letting us see the microfilm, K. Fukuda and T. Sugawa for access to the microfilm, R. Uemura and Y. Kobayashi for help preparing analyses of sunspot area, Y. Mizuno for technical supports for trace copy, M. Morishita and M. Woods for grammatical review, Y. Sakai for his advices on Kunitomo's telescope, SILSO for providing the international sunspot number, and E. W. Cliver for scientific comments and preliminary review to our study. This work was supported by JSPS KAKENHI Grant Numbers JP16K17671 and JP15H05814. This work is partially supported by MEXT/JSPS KAKENHI Grant Number 15H05816.

**Disclosure of Potential Conflict of Interest:**

The authors declare that they have no conflict of interest.

Table 1 RGC catalog for S and its comparison to previous studies: K32 (Kanda 1932), T98 (Tomita *et al.*, 1998), Y35 (Yamamoto, 1935). Yamamoto and we distinguish morning and afternoon. In this table, month, day, and time are given in local mean time at the Kunitomo's observational site in Japan. The symbol "g" notes sunspot group number. The entry with "-1" means there were no observations or data.

| ID | YEAR | MONTH | DAY | AM | PM | K32 | T98 | Y35 | | Ours | |
|---|---|---|---|---|---|---|---|---|---|---|---|
| | | | | | | g | g | g(AM) | g(PM) | g(AM) | g(PM) |
| 03_A* | 1835 | 2 | 3 | 8 | N/A | 1 | 1 | 1 | -1 | 1 | -1 |
| 03_B* | 1835 | 2 | 6 | 8 | N/A | 1 | 1 | 1 | -1 | 1 | -1 |
| 03_C* | 1835 | 2 | 7 | 8 | N/A | 1 | 1 | 1 | -1 | 1 | -1 |
| 03_D* | 1835 | 2 | 8 | 8 | N/A | 1 | 1 | 1 | -1 | 1 | -1 |
| 04_A* | 1835 | 2 | 9 | 8 | N/A | 1 | 1 | 1 | -1 | 1 | -1 |
| N/A | 1835 | 2 | 10 | N/A | N/A | -1 | 2 | 0 | -1 | -1 | -1 |
| 04_B* | 1835 | 3 | 2 | 11 | N/A | 1 | -1 | 1 | -1 | 2 | -1 |
| 04_C* | 1835 | 3 | 3 | N/A | N/A | -1 | -1 | -1 | -1 | -1 | -1 |
| 04_D* | 1835 | 3 | 4 | 11 | 14 | 1 | 2 | 1 | 1 | 2 | 2 |
| 05_A* | 1835 | 3 | 5 | 11 | N/A | 1 | 2 | 1 | -1 | 2 | -1 |
| 05_B* | 1835 | 3 | 6 | N/A | 14 | 1 | 3 | -1 | 1 | -1 | 3 |
| 05_C* | 1835 | 3 | 8 | 10 | N/A | 1 | 2 | 1 | -1 | 2 | -1 |
| 05_D* | 1835 | 4 | 3 | 10 | 14 | 1 | 1 | 1(2) | 1 | 1 | 1 |
| 05_E* | 1835 | 4 | 5 | N/A | 14 | 2 | 2 | -1 | 2 | -1 | 2 |
| 06_A* | 1835 | 4 | 6 | 10 | 14 | 2 | 2 | 2 | 2 | 2 | 2 |
| 06_B* | 1835 | 4 | 8 | N/A | 14 | 2 | 2 | -1 | 2 | -1 | 2 |
| 06_C* | 1835 | 4 | 9 | N/A | 14 | 3 | 4 | -1 | 3(3) | -1 | 4 |
| 06_D* | 1835 | 4 | 11 | 9 | 14 | 3 | 4 | 3(4) | 2(2) | 3 | 2 |
| 06_E* | 1835 | 4 | 12 | 8 | N/A | 3 | 4 | 3 | -1 | 3 | -1 |
| 07_A* | 1835 | 4 | 14 | 9 | N/A | 2 | 2 | 2 | -1 | 2 | -1 |
| 07_B* | 1835 | 4 | 16 | 8 | 16 | 2 | 2 | 2(2) | 2(2) | 2 | 2 |
| 07_C* | 1835 | 4 | 17 | 9 | N/A | 2 | 2 | 2 | -1 | 2 | -1 |
| 07_D* | 1835 | 4 | 18 | 9 | N/A | 2 | 2 | 2 | -1 | 2 | -1 |
| 07_E* | 1835 | 4 | 19 | 8 | N/A | 1 | 1 | 2 | -1 | 1 | -1 |
| 08_A* | 1835 | 4 | 20 | 8 | N/A | 1 | 1 | 1 | -1 | 1 | -1 |
| 08_B* | 1835 | 4 | 21 | 8 | N/A | 1 | 1 | 1 | -1 | 1 | -1 |
| 08_C* | 1835 | 4 | 22 | 8 | N/A | 2 | 2 | 2 | -1 | 2 | -1 |
| 08_D* | 1835 | 4 | 24 | 8 | N/A | 2 | 3 | 2 | -1 | 3 | -1 |
| 08_E* | 1835 | 4 | 25 | 8 | N/A | 1 | 2 | 1 | -1 | 2 | -1 |
| 09_A* | 1835 | 4 | 26 | 7 | N/A | 1 | 2 | 1 | -1 | 2 | -1 |
| 09_B* | 1835 | 4 | 27 | 10 | N/A | 1 | 2 | 1 | -1 | 2 | -1 |
| 09_C* | 1835 | 4 | 28 | 9 | N/A | 2 | 3 | 2 | -1 | 3 | -1 |
| 09_D* | 1835 | 5 | 2 | 8 | N/A | 3 | 3 | 3 | -1 | 3 | -1 |
| 09_E* | 1835 | 5 | 3 | 8 | N/A | 3 | 3 | 3 | -1 | 3 | -1 |
| 10_A* | 1835 | 5 | 4 | 8 | N/A | 4 | 4 | 4 | -1 | 4 | -1 |
| 10_B* | 1835 | 5 | 6 | 9 | N/A | 4 | 5 | 4 | -1 | 5 | -1 |
| 10_C* | 1835 | 5 | 7 | 8 | 14 | 5 | 5 | 5 | 5 | 5 | 5 |
| 10_D* | 1835 | 5 | 8 | 9 | N/A | 5 | 7 | 6 | -1 | 6 | -1 |
| 10_E* | 1835 | 5 | 9 | 9 | N/A | 3 | 5 | 5 | -1 | 4 | -1 |
| 11_A* | 1835 | 5 | 11 | 9 | N/A | 2 | -1 | 2 | -1 | 2 | -1 |
| 11_B* | 1835 | 5 | 12 | 9 | N/A | 2 | -1 | 2 | -1 | 2 | -1 |
| 11_C* | 1835 | 5 | 13 | 9 | N/A | 2 | -1 | 2 | -1 | 2 | -1 |
| 11_D* | 1835 | 5 | 17 | 9 | N/A | 2 | -1 | 2 | -1 | 2 | -1 |
| 11_E* | 1835 | 5 | 19 | 9 | N/A | 2 | -1 | 2 | -1 | 2 | -1 |
| 12_A* | 1835 | 5 | 22 | 8 | N/A | 2 | 2 | 2 | -1 | 2 | -1 |
| 12_B* | 1835 | 5 | 25 | 14 | N/A | 1 | 1 | -1 | 1 | -1 | 1 |
| 12_C* | 1835 | 5 | 26 | 8 | N/A | 1 | 1 | 1 | -1 | 1 | -1 |
| 12_D* | 1835 | 5 | 30 | 8 | 17 | 1 | 1 | 1 | 1 | 1 | 1 |
| 12_E* | 1835 | 5 | 31 | 8 | 16 | 1 | 1 | -1 | 1 | 1 | 1 |
| 13_A* | 1835 | 6 | 4 | 8 | 17 | 2 | 2 | 1 | 2 | 2 | 2 |
| 13_B* | 1835 | 6 | 5 | 8 | N/A | 2 | 2 | 2 | -1 | 2 | -1 |
| 13_C* | 1835 | 6 | 7 | 9 | 16 | 3 | 3 | 2 | 3 | 3 | 3 |
| 13_D* | 1835 | 6 | 10 | 11 | N/A | 2 | 2 | 3 | -1 | 2 | -1 |
| 13_E* | 1835 | 6 | 11 | 10 | 15 | 2 | 2 | 2 | 2 | 2 | 2 |
| 14_A* | 1835 | 6 | 12 | 9 | 16 | 2 | 2 | 2 | 2 | 2 | 2 |
| 14_B* | 1835 | 6 | 13 | 8 | 16 | 2 | 2 | 2 | 2 | 2 | 2 |
| 14_C* | 1835 | 6 | 14 | 8 | N/A | 2 | 2 | 2 | -1 | 2 | -1 |
| 14_D* | 1835 | 7 | 12 | N/A | 15 | 3 | 3 | -1 | 3 | -1 | 3 |
| 14_E* | 1835 | 7 | 13 | 9 | N/A | 3 | 5 | 3 | -1 | 4 | -1 |
| 15_A* | 1835 | 7 | 14 | 9 | N/A | 1 | 3 | 1 | -1 | 2 | -1 |
| 15_B* | 1835 | 7 | 15 | 9 | N/A | 2 | 5 | 2 | -1 | 3 | -1 |
| 15_C* | 1835 | 7 | 17 | 8 | 16 | 1 | 3 | 1 | 1 | 2 | 2 |
| 15_D* | 1835 | 7 | 18 | 8 | N/A | 2 | 4 | 3 | -1 | 3 | -1 |
| 15_E* | 1835 | 7 | 19 | 9 | N/A | 2 | 4 | 3 | -1 | 3 | -1 |
| 16_A* | 1835 | 7 | 20 | 8 | 16 | 1 | 3 | 2 | 1 | 1 | 1 |
| 16_B* | 1835 | 7 | 21 | 8 | N/A | 1 | 2 | 2 | -1 | 1 | -1 |
| 16_C* | 1835 | 7 | 26 | 9 | 17 | 3 | 3 | 3 | 3 | 3 | 3 |
| 16_D* | 1835 | 7 | 27 | 8 | 16 | 3 | 4 | 3 | 2 | 3 | 1 |
| 17_A* | 1835 | 8 | 6 | 8 | N/A | 1 | 2 | 1 | -1 | 1 | -1 |
| 17_B* | 1835 | 8 | 7 | N/A | 16 | 1 | 2 | -1 | 1 | -1 | 2 |
| 17_C* | 1835 | 8 | 8 | 8 | N/A | 1 | 2 | 1 | -1 | 2 | -1 |
| 17_D* | 1835 | 8 | 9 | 8 | 16 | 2 | 3 | 1 | 2 | 3 | 3 |
| 17_E* | 1835 | 8 | 10 | 8 | 16 | 2 | 3 | 2 | 2 | 3 | 3 |
| 18_A* | 1835 | 8 | 11 | 8 | 16 | 1 | 2 | 1 | 1 | 2 | 2 |
| 18_B* | 1835 | 8 | 12 | 8 | 16 | 2 | 3 | 2 | 2 | 3 | 2 |
| 18_C* | 1835 | 8 | 13 | 8 | 16 | 2 | 3 | 1 | 2 | 1 | 3 |
| 18_D* | 1835 | 8 | 14 | 8 | 16 | 2 | 4 | 2 | 2 | 3 | 2 |
| 18_E* | 1835 | 8 | 18 | 8 | 16 | 1 | 1 | 1 | 1 | 1 | 1 |
| 19_A* | 1835 | 8 | 19 | 8 | N/A | 2 | 3 | 2 | -1 | 2 | -1 |
| 19_B* | 1835 | 8 | 20 | 10 | 16 | 3 | 4 | 3 | 4 | 4 | 4 |
| 19_C* | 1835 | 8 | 21 | 8 | 16 | 3 | 4 | 4 | 3 | 4 | 3 |
| 19_D* | 1835 | 8 | 22 | 7 | 16 | 1 | 2 | 2 | 2 | 2 | 2 |
| 19_E* | 1835 | 8 | 23 | 7 | 16 | 2 | 3 | 3 | 3 | 3 | 3 |
| 20_A* | 1835 | 8 | 24 | 8 | 16 | 2 | 2 | 2 | 2 | 2 | 2 |
| 20_B* | 1835 | 8 | 25 | 7 | 16 | 2 | 2 | 2 | 2 | 2 | 2 |
| 20_C* | 1835 | 8 | 26 | 6 | N/A | 3 | 3 | 3 | -1 | 3 | -1 |
| 20_D* | 1835 | 8 | 27 | 6 | 16 | 3 | 3 | 3 | 3 | 3 | 3 |
| 20_E* | 1835 | 8 | 28 | 8 | 18 | 2 | 3 | 3 | 2 | 3 | 2 |
| 21_A* | 1835 | 8 | 31 | N/A | 16 | 2 | 2 | -1 | 2 | -1 | 2 |
| 21_B* | 1835 | 9 | 1 | 8 | 16 | 2 | 2 | 2 | 2 | 3 | 3 |
| 21_C* | 1835 | 9 | 2 | 9 | N/A | 2 | 2 | 2(2) | -1 | 3 | -1 |
| 21_D* | 1835 | 9 | 4 | 8 | 16 | 2 | 3 | 2(2) | 2(2) | 3 | 3 |
| 21_E* | 1835 | 9 | 5 | 8 | N/A | 2 | 3 | 2(2) | 2(2) | 3 | -1 |
| 22_A* | 1835 | 9 | 6 | 8 | 16 | 5 | 6 | 5(5) | 4(4) | 6 | 4 |
| 22_B* | 1835 | 9 | 7 | 8 | 16 | 3 | 4 | 3(3) | 3(3) | 4 | 4 |
| 22_C* | 1835 | 9 | 8 | 8 | 16 | 4 | 5 | 4(4) | 4(4) | 5 | 5 |
| 22_D* | 1835 | 9 | 9 | 10 | 16 | 4 | 5 | 4(4) | 4(4) | 5 | 5 |
| 22_E* | 1835 | 9 | 10 | 8 | 15 | 4 | 5 | 4(4) | 3(3) | 5 | 3 |
| 23_A* | 1835 | 9 | 11 | 8 | 16 | 3 | 4 | 3(2) | 4(4) | 3 | 4 |
| 23_B* | 1835 | 9 | 12 | 8 | N/A | 4 | 5 | 4(4) | 4(4) | 4 | 4 |
| 23_C* | 1835 | 9 | 13 | 8 | N/A | 4 | 5 | 4(4) | -1 | 6 | -1 |
| 23_D* | 1835 | 9 | 15 | 8 | N/A | 3 | 6 | 5(6) | -1 | 5 | -1 |
| 23_E* | 1835 | 9 | 16 | N/A | N/A | 3 | 7 | 5(6) | 3(6) | 5 | 6 |
| 24_A* | 1835 | 9 | 17 | 8 | N/A | 3 | 5 | 5(6) | -1 | 5 | -1 |
| 24_B* | 1835 | 9 | 20 | 8 | N/A | 3 | 4 | 3(4) | -1 | 3 | -1 |
| 24_C* | 1835 | 9 | 21 | N/A | N/A | 3 | 4 | 4(4) | -1 | 3 | -1 |
| 24_D* | 1835 | 9 | 22 | 8 | 17 | 3 | 7 | 3(6) | 3(5) | 6 | 6 |
| 24_E* | 1835 | 9 | 24 | 8 | 16 | 3 | 5 | 3(5) | 3(6) | 5 | 4 |
| 25_A* | 1835 | 9 | 25 | 8 | N/A | 3 | 5 | 5 | -1 | 4 | -1 |
| 25_B* | 1835 | 9 | 28 | 9 | N/A | 2 | 3 | 3 | -1 | 3 | -1 |
| 25_C* | 1835 | 9 | 29 | 8 | N/A | 2 | 3 | 2 | -1 | 3 | -1 |
| 25_D* | 1835 | 9 | 30 | 8 | N/A | 2 | 3 | 2 | -1 | 3 | -1 |
| 25_E* | 1835 | 10 | 2 | 8 | 16 | 2 | 3 | 2 | 2 | 3 | 3 |
| 26_A* | 1835 | 10 | 3 | 6 | N/A | 1 | 1 | 1 | -1 | 1 | -1 |
| 26_B* | 1835 | 10 | 4 | 8 | 14 | 1 | 1 | 1 | 1 | 1 | 1 |
| 26_C* | 1835 | 10 | 8 | 8 | 16 | 1 | 1 | 1 | 1 | 1 | 1 |
| 26_D* | 1835 | 10 | 10 | 8 | N/A | 2 | 2 | 2 | -1 | 2 | -1 |
| 26_E* | 1835 | 10 | 11 | 8 | N/A | 2 | 2 | 2 | -1 | 2 | -1 |
| 27_A* | 1835 | 10 | 13 | 8 | N/A | 3 | 3 | 3 | -1 | 3 | -1 |
| 27_B* | 1835 | 10 | 14 | 8 | 16 | 3 | 3 | 3 | 3 | 3 | 3 |
| 27_C* | 1835 | 11 | 8 | 8 | N/A | 3 | 3 | 3 | -1 | 3 | -1 |
| 27_D* | 1835 | 11 | 13 | 8 | N/A | 5 | 5 | 4 | -1 | 4 | -1 |
| 27_E* | 1835 | 11 | 14 | 8 | 16 | 4 | 7 | 4 | 4 | 4 | 6 |
| 28_A* | 1835 | 11 | 15 | 8 | 16 | 4 | 7 | 5 | 5 | 6 | 5 |
| 28_B* | 1835 | 11 | 17 | 8 | 16 | 4 | 7 | 7 | 6 | 6 | 6 |
| 28_C* | 1835 | 11 | 23 | 8 | N/A | 1 | 1 | 1 | -1 | 1 | -1 |
| 28_D* | 1835 | 11 | 26 | 8 | 16 | 2 | 2 | 1 | 2 | 1 | 2 |
| 28_E* | 1835 | 11 | 27 | 8 | N/A | 2 | 3 | 2 | -1 | 2 | -1 |
| 29_A* | 1835 | 11 | 28 | 8 | N/A | 3 | 4 | 3 | -1 | 4 | -1 |
| 29_B* | 1835 | 11 | 29 | 8 | 16 | 3 | 4 | 4 | 4 | 4 | 4 |
| 29_C* | 1835 | 11 | 30 | N/A | 16 | 5 | 5 | -1 | 5 | -1 | 5 |
| 29_D* | 1835 | 12 | 2 | N/A | 16 | 5 | 5 | -1 | 5 | -1 | 5 |
| 29_E* | 1835 | 12 | 3 | N/A | 14 | 6 | 6 | -1 | 6 | -1 | 6 |
| 30_A* | 1835 | 12 | 4 | 8 | 16 | 5 | 5 | 5 | 5 | 5 | 5 |
| 30_B* | 1835 | 12 | 5 | 8 | N/A | 4 | 4 | 4 | -1 | 4 | -1 |
| 30_C* | 1835 | 12 | 6 | N/A | 16 | 3 | 5 | -1 | 4 | -1 | 3 |
| 30_D* | 1835 | 12 | 12 | N/A | N/A | 3 | 4 | 4 | -1 | 4 | -1 |
| 31_A* | 1835 | 12 | 17 | 8 | N/A | 5 | 5 | 5 | -1 | 7 | -1 |
| 31_B* | 1835 | 12 | 19 | 8 | N/A | 3 | 4 | 3 | -1 | 4 | -1 |
| 31_C* | 1835 | 12 | 20 | 8 | N/A | 4 | 4 | 5 | -1 | 4 | -1 |
| 31_D* | 1836 | 1 | 1 | 8 | N/A | 1 | 2 | 1 | -1 | 1 | -1 |
| 31_E* | 1836 | 1 | 3 | 8 | N/A | 1 | 2 | 1 | -1 | 1 | -1 |
| 32_A* | 1836 | 1 | 5 | 8 | N/A | 2 | 3 | 2 | -1 | 2 | -1 |
| 32_B* | 1835 | 12 | 8 | 8 | N/A | -1 | -1 | -1 | -1 | 4 | -1 |
| N/A | 1836 | 1 | 26 | N/A | N/A | 4 | 5 | 4 | -1 | -1 | -1 |
| 32_C* | 1836 | 2 | 20 | 8 | 14 | 3 | 3 | 3 | 3 | 3 | 3 |
| 32_D* | 1836 | 2 | 21 | 8 | 16 | 3 | 3 | 3 | 3 | 3 | 3 |
| 33_A* | 1836 | 2 | 23 | N/A | 14 | 3 | 3 | -1 | 3 | -1 | 3 |
| 33_B* | 1836 | 2 | 24 | 8 | N/A | 4 | 5 | 4 | -1 | 5 | -1 |
| 33_C* | 1836 | 2 | 25 | 8 | N/A | 5 | 5 | 5 | -1 | 5 | -1 |
| 33_D* | 1836 | 2 | 26 | 8 | N/A | 5 | 5 | 5 | -1 | 5 | -1 |
| 33_E* | 1836 | 2 | 28 | 8 | N/A | 6 | 6 | 6 | -1 | 6 | -1 |
| 34_A* | 1836 | 3 | 5 | 9 | N/A | 3 | 5 | 5 | -1 | 4 | -1 |
| 34_B* | 1836 | 3 | 7 | 9 | N/A | 4 | 5 | 4 | -1 | 5 | -1 |
| 34_C* | 1836 | 3 | 11 | 8 | N/A | 3 | 5 | 5 | -1 | 3 | -1 |
| 34_D* | 1836 | 3 | 13 | 10 | N/A | 5 | 7 | 6 | -1 | 5 | -1 |
| 34_E* | 1836 | 3 | 19 | 8 | N/A | 3 | 5 | 3 | -1 | 4 | -1 |
| 35_A* | 1836 | 3 | 22 | N/A | 16 | 2 | 2 | -1 | 2 | -1 | 2 |
| 35_B* | 1836 | 3 | 23 | 8 | 16 | 3 | 3 | 3 | -1 | 3 | 3 |
| 35_C* | 1836 | 3 | 24 | 8 | N/A | 3 | 3 | 3 | 3 | 3 | -1 |

Table 2 RGC catalog for P and I. When the solar disk does not show any sunspot, -1 is filled in.

| ID(P) | Corresponding S | DATE | g |
|---|---|---|---|
| P_A0 | 21_C0 | 1835/09/02 | 4 |
| P_A1 | 21_D0 | 1835/09/04 | 3 |
| P_B0 | 21_D0 | 1835/09/04 | 3 |
| P_B1 | 21_D1 | 1835/09/04 | 5 |
| P_C0 | 21_E0 | 1835/09/05 | 4 |
| P_C1 | 21_E1 | 1835/09/05 | 3 |
| P_D0 | 22_A0 | 1835/09/06 | 7 |
| P_D1 | 22_A1 | 1835/09/06 | 7 |
| P_E0 | 22_B0 | 1835/09/07 | 9 |
| P_E1 | 22_B1 | 1835/09/07 | 4 |
| P_F0 | 22_C0 | 1835/09/08 | 6 |
| P_F1 | 22_C1 | 1835/09/08 | 6 |
| P_G0 | 22_D0 | 1835/09/09 | 5 |
| P_G1 | 22_D1 | 1835/09/09 | 5 |
| P_H0 | 22_E0 | 1835/09/10 | 7 |
| P_H1 | 22_E1 | 1835/09/10 | 4 |
| P_I0 | 23_A0 | 1835/09/11 | 4 |
| P_I1 | 23_A1 | 1835/09/11 | 5 |
| P_J0 | 23_B0 | 1835/09/12 | 4 |
| P_J1 | 23_B1 | 1835/09/12 | 4 |
| P_K0 | 23_C0 | 1835/09/13 | 6 |
| P_K1 | 23_C1 | 1835/09/13 | −1 |
| P_L0 | 23_D0 | 1835/09/15 | 5 |
| P_L1 | 23_D1 | 1835/09/15 | −1 |
| P_M0 | 23_E0 | 1835/09/16 | 7 |
| P_M1 | 23_E1 | 1835/09/16 | 8 |
| P_N0 | 24_A0 | 1835/09/17 | 5 |
| P_N1 | 24_A1 | 1835/09/17 | −1 |
| P_O0 | 24_B0 | 1835/09/20 | 5 |
| P_O1 | 24_C0 | 1835/09/21 | 3 |
| P_P0 | 24_D0 | 1835/09/22 | 6 |
| P_P1 | 24_D1 | 1835/09/22 | 6 |
| P_Q0 | 24_E0 | 1835/09/24 | 5 |
| P_Q1 | 24_E1 | 1835/09/24 | 5 |
| I | | 1836/09/25 | 7 |